\documentclass[11pt,a4paper,onecolumn,twoside]{rho}
\usepackage[english]{babel}
\usepackage{amsmath,amssymb,physics}

\usepackage{hyperref}
\usepackage{comment}
\newcommand{\eq}[1]{Eq.~(\ref{#1})} %
\usepackage{acro}

\DeclareAcronym{vqe}{
  short = VQE,
  long  = Variational Quantum Eigensolver,
}

\DeclareAcronym{oo-vqe}{
  short = oo-VQE,
  long  = orbital-optimized Variational Quantum Eigensolver,
}

\DeclareAcronym{pp-tUPS}{
  short = pp-tUPS,
  long  = perfect-pairing tiled Unitary Product State,
}

\DeclareAcronym{bfgs}{
  short = BFGS,
  long  = Broyden–Fletcher–Goldfarb–Shanno,
}

\DeclareAcronym{PES}{
  short = PES,
  long  = potential energy surface,
}

\DeclareAcronym{SI}{
  short = SI,
  long  = supplementary information,
}

\DeclareAcronym{QPU}{
  short = QPU,
  long  = quantum processing units,
}

\DeclareAcronym{qesem}{
    short = QESEM,
    long = Quantum Error Suppression and Error Mitigation,
}

\DeclareAcronym{ec}{
    short = EC,
    long = Error Correction,
}

\DeclareAcronym{qem}{
    short = QEM,
    long = Quantum Error Mitigation,
}

\DeclareAcronym{qec}{
    short = QEC,
    long = Quantum Error Correction,
}

\DeclareAcronym{std}{
    short = std,
    long = standard deviation,
}

\DeclareAcronym{NISQ}{
    short = NISQ,
    long = Noisy Intermediate-Scale Quantum,
}

\DeclareAcronym{zne}{
    short = ZNE,
    long = zero-noise extrapolation,
}

\DeclareAcronym{pec}{
    short = PEC,
    long = probabilistic error cancellation,
}

\doctype{Research Article}
\title{Implementing QESEM's High-Accuracy Error Mitigation on a Quantum Computer: a Water Potential Energy Surface Study}

\author[{\textasteriskcentered},a]{Renato Olarte Hernandez}
\author[b]{Emanuele Rossi}
\author[b]{Sonia Coriani}
\author[b,c]{Karl Michael Ziems}
\author[d]{Erik Kjellgren}
\author[d]{Jacob Kongsted}
\author[e]{Tali Shnaider}
\author[a]{Stephan P. A. Sauer}

\affil[a]{Department of Chemistry, University of
Copenhagen, DK-2100 Copenhagen, Denmark.}
\affil[b]{DTU Chemistry - Department of Chemistry, Technical
University of Denmark, DK-2800 Kongens Lyngby,
Denmark.}
\affil[c]{School of Chemistry, University of Southampton, Highfield, Southampton SO17 1BJ, United Kingdom}
\affil[d]{Department of Physics, Chemistry and Pharmacy, University of Southern Denmark, DK-5230 Odense, Denmark.}
\affil[e]{Qedma Quantum Computing, Tel Aviv, Israel}

\corres{\textsuperscript{\textasteriskcentered}Corresponding author: \href{mailto:author.one@institute.org}{rehe@chem.ku.dk}.}

\journalname{This research article was typeset in LaTeX with the rho class.}
\journal{København Universitet}
\theday{\today}

\begin{abstract}

Quantum error mitigation (QEM) is essential for extracting chemically accurate results from near-term quantum hardware. Many widely used QEM methods rely on uncontrolled heuristics whose bias depends on the specific circuit and noise realization. In this work, we employ QESEM---a characterization-based, unbiased quasi-probabilistic mitigation method---on IBM's Aachen quantum processor to compute the ground-state potential energy surface (PES) of the symmetrically-stretched water molecule. We consider a classically-optimized, single-layer perfect-pairing tiled unitary product state (pp-tUPS) ansatz. We map this ansatz to an 8-qubit register corresponding to a (4,4) active space and the STO-3G basis set. Compared to the statevector reference, we find that raw QPU results typically overestimate the ground-state energy by around 500~mHa across the scanned geometries. On the other hand, QESEM-mitigated results fall approximately within 100~mHa, 30~mHa, or 
``chemical accuracy'' ($\sim$1.5~mHa), 
depending on the target precision. We benchmark QESEM at both loose (0.1~Ha) and tight (0.01~Ha) precision targets, evaluating both individual and merged batches of runs. As the precision target is tightened, the accuracy improves systematically, matching or exceeding results reported in the literature. We further quantify the sampling cost of these results, reporting the number of shots required at each precision level. Our results show how, with the current levels of hardware error, reaching the highest accuracies demands substantial QPU time. The results demonstrate that the characterization-based, unbiased error mitigation provided by QESEM allows to measure quantitatively meaningful potential energy surfaces on current quantum hardware. Concurrently, they highlight the sampling overhead associated with QEM, which remains a central bottleneck en route to larger chemical problems and higher precision.

\end{abstract}

\begin{document}

    \maketitle
    \thispagestyle{firststyle}


\section{Introduction}
Achieving \textit{quantum advantage} \cite{lanes2025framework,Google_QA_2025,Liu_QA_2025} crucially relies on addressing noise sources such as gate infidelities, decoherence, and readout errors characterizing the operations of quantum hardware \cite{aharonov2025importance,lanes2025framework,preskill2018quantum}. Left unaddressed, hardware errors accumulate over the course of a quantum computation, posing severe limitations on the circuit depth and the accuracy of the outputs. While full \ac{qec} represents the definitive step towards large-scale quantum advantage, its requirements for large qubit numbers, high-fidelity operations, and adaptive quantum operations make this approach unfeasible on currently available \ac{NISQ} devices \cite{shor1996fault,aharonov1997fault}.

\ac{qem} represents a valid approach to suppress errors on \ac{NISQ} devices and possibly provide finite quantum advantage in the near future \cite{aharonov2025importance}. \ac{qem} methods do not rely on adaptive quantum operations and require a limited qubit overhead; instead, they build upon multiple noisy executions of the quantum circuit and classical post-processing to estimate the error-free result of the circuit measurement \cite{cao2021nisq,suzuki2022quantum,cai2023quantum}. A variety of \ac{qem} protocols---each trading a larger sampling overhead for higher accuracy---have been developed and tested extensively on real quantum processors \cite{khan2024error,Roffe_2019,suzuki2022quantum,google2025quantum}. Techniques such as dynamical decoupling \cite{khodjasteh2005fault,souza2011robust} operate directly at the hardware or pulse level, while approaches like twirled readout error extraction \cite{belaloui2026improving} and readout error mitigation \cite{bravyi2021mitigating} act at the measurement or post-processing stage. A family of approaches including \ac{zne} \cite{temme2017error,li2017efficient}, \ac{pec} \cite{temme2017error,endo2018practical}, Clifford data regression \cite{czarnik2021error}, and Ansatz-based gate and readout error mitigation \cite{ziems2025understanding,rasmussen2026jctc} reconstructs or extrapolates the noiseless result from an ensemble of noisy circuit executions. Many of these methods generally offer no guarantee on the resulting output error \cite{cai2021multi}, as they rely on uncontrolled heuristics based on the specific hardware noise and circuit being run.

This lack of guarantees is particularly limiting for quantum chemistry applications, where optimizing the ansatz to \textit{chemical accuracy} (typically on the order of $1$~kcal/mol from the true experimental value, and as an extension herein, from the ideal theoretical reference) is required to reliably predict reaction energies and compete with established classical approximation methods \cite{mcardle2020quantum}. As circuit sizes grow toward regimes that are increasingly hard to verify against classical simulation, it becomes essential to use mitigation techniques which can systematically bound and, ideally, eliminate the bias error of quantum estimations. Quasi-probabilistic methods such as \ac{pec} \cite{temme2017error,endo2018practical} are unbiased in principle (as long as the noise model is characterized accurately): in the limit of infinitely many circuit samples, their estimates converge to the exact noiseless value. However, standard \ac{pec} incurs a prohibitive sampling overhead that has restricted its experimental demonstration to comparatively small circuits \cite{van2023probabilistic,ferracin2024efficiently}. 
In this work, we use the recently developed unbiased \ac{qesem} software \cite{aharonov2025reliable}, which substantially reduces the runtime overhead by a combination of noise-aware circuit transpilation, supporting direct mitigation of non-Clifford fractional gates without needing to transpile them to Clifford gates which greatly reduces the circuit volume, active-volume identification, optimized allocation of \ac{QPU} time across different measurement bases for complex observables. The fact that QESEM supports mitigation of circuits which contains both fractional and Clifford gates makes it well suited to the quantum chemistry ansatz circuits considered here. 

We benchmark the capabilities of \ac{qesem} in the context of the measurement of the ground-state \ac{PES} associated with the symmetric stretching mode of the water molecule. Potential energy surfaces are a central object in quantum chemistry, underlying the prediction of equilibrium geometries, reaction pathways, and spectroscopic properties of molecules \cite{atkins2023atkins}. The accurate reconstruction of a \ac{PES} requires computing the ground-state energy at many nuclear configurations to chemical accuracy, which places heavy demands both on the electronic structure method itself and on controlling statistical and systematic errors at every point on the surface. On quantum hardware, this requirement is compounded by hardware noise. If left unmitigated, hardware noise distorts the shape of the surface and introduces false features that are not merely energy offsets but can qualitatively affect derived properties such as equilibrium bond lengths or force constants. 
The water molecule is a particularly demanding benchmark: despite its modest size, determining its \ac{PES} entails the description of a non-trivial degree of electron correlation across the whole range of scanned geometries. This makes the geometry-dependent energy measurements in water noticeably more sensitive to residual mitigation bias than in smaller, more commonly simulated diatomic systems \cite{hernandez2026analytical,zou2025multireference}.

We tackle the measurement of the \ac{PES} on quantum hardware using the \ac{pp-tUPS} ansatz \cite{Burton_2024}. This hardware-efficient, chemically motivated circuit has shown promise for extending quantum chemistry simulations beyond ground-state energies to derivative properties on real hardware \cite{hernandez2026analytical}. We established an exact reference for the measurements on quantum hardware by optimizing the pp-tUPS ansatz using the \ac{vqe} protocol \cite{peruzzo2014variational,mcclean2016theory} on a statevector simulator. In particular, we restricted the simulation to an active space and applied the \ac{oo-vqe} method \cite{Mizukami_2020, Sokolov_2020} to optimize the orbitals alongside the circuit parameters. We measured the energy expectation values of the optimized pp-tUPS circuits at each molecular geometry on IBM's Aachen quantum processor adopting \ac{qesem} to perform \ac{qem}. This combination targets the specific challenge outlined above: an unbiased, characterization-based mitigation scheme applied to a chemically motivated ansatz, benchmarked on a molecule whose \ac{PES} is demanding enough to expose the limits of heuristic mitigation approaches.

This article is organized as follows: the Theory section (\ref{theory}) describes the fundamental theory behind the wavefunction parameterization, quantum circuit, energy calculation, and the \ac{qesem} method. The Computational Methods section (\ref{comp_meth}) describes the level of theory, the tools used for result analysis, and the computational details of the simulations. The Results and Discussion section (\ref{results_disc}) presents the results for the water molecule, compared against CASSCF and ideal \ac{QPU} simulations. Conclusions are drawn in Conclusions and Outlook (\ref{conclusions}).

\section{Theory}\label{theory}
In the following, we present the theoretical background underlying the implementation. In Sec.~\ref{sec:H_active_space} we discuss the definition of the Hamiltonian and the wave function ansatz within the active space approximation. In Sec.~\ref{sec:oo-VQE} we briefly discuss the (parameter and orbital) optimization of the wave function ansatz and the estimation of its energy on quantum hardware. In Sec.~\ref{sec:tUPS-ansatz} we discuss the properties of the \ac{pp-tUPS} ansatz \cite{Burton_2024} we utilized in this work. Finally, in Sec.~\ref{sec:QESEM-em} we discuss the \ac{qesem} error mitigation framework.

\subsection{Hamiltonian and wave function within the active space approximation}\label{sec:H_active_space}
We consider the Hamiltonian $\hat{\text{H}}(\boldsymbol{\kappa})$ in the Born-Oppenheimer approximation, which in second quantization reads

\begin{equation}\label{eqn:electronic_H}
    \hat{\text{H}}(\boldsymbol{\kappa}) = \sum_{pq}h_{pq}(\boldsymbol{\kappa})\hat{\text{E}}_{pq} + \frac{1}{2}\sum_{pqrs}g_{pqrs}(\boldsymbol{\kappa})\hat e_{pqrs} + h_{\text{nuc}}~
    \equiv \hat{\text{H}}_\textrm{el}(\boldsymbol{\kappa}) +
    h_{\text{nuc}}
\end{equation}
Here, $\hat{\text{E}}_{pq} = \hat{a}_{p\alpha}^{\dagger}\hat{a}_{q\alpha} + \hat{a}_{p\beta}^{\dagger}\hat{a}_{q\beta}$ are the singlet one-electron excitation operators (where $\alpha$ and $\beta$ indicate the spin orientation) and $h_{pq}(\boldsymbol{\kappa})$ are the one-electron integrals; $\hat e_{pqrs} =  \hat{\text{E}}_{pq} \hat{\text{E}}_{rs} - \delta_{qr}\hat{\text{E}}_{ps}$ are the singlet two-electrons excitation operators and $g_{pqrs}(\boldsymbol{\kappa})$  the two-electron integrals. The indices $p$, $q$, $r$, $s$ label general molecular orbitals;  $\boldsymbol{\kappa}$ corresponds to the vector of orbital rotation parameters defining the molecular orbitals. The first two terms in \eq{eqn:electronic_H} correspond to the electronic Hamiltonian $\hat{\text{H}}_{\text{el}}(\boldsymbol{\kappa})$, while $h_{\text{nuc}}$ is the nuclear repulsion energy (which is a constant term depending on the nuclear geometry).

In the active space approximation, the wave function can be written as

\begin{equation}\label{eqn:AS_appr}
    \ket{\Psi(\boldsymbol{\theta})} = \ket{\Psi}_{\text{I}}\otimes\ket{\Psi(\boldsymbol{\theta})}_{\text{A}}\otimes\ket{\Psi}_{\text{V}},
\end{equation}
where $\ket{\Psi}_{\text{I}}$ corresponds to the \textit{inactive} orbitals (doubly occupied in all configurations and indicated by indices $i, j, k, l$), $\ket*{\Psi(\boldsymbol{\theta})}_{\text{A}}$ corresponds to the \textit{active} orbitals (with no restrictions on their occupations and indicated by indices $u, v, x, y$), and $\ket{\Psi}_{\text{V}}$ corresponds to the \textit{virtual} orbitals (unoccupied in all configurations and indicated by indices $a, b, c, d$). The vector $\boldsymbol{\theta}$ collects the ansatz parameters, which, as can be seen in \eq{eqn:AS_appr}, parametrize solely the active part of the wave function ansatz. 

The evaluation of the electronic energy relies on splitting $\hat{\text{H}}_{\text{el}}(\boldsymbol{\kappa})$ in active and inactive parts, with the simulation on the quantum computer being limited to the sole active orbitals. Within this approach, the inactive electrons are treated classically  and act as a mean-field environment for the active electrons. The active part of $\hat{\text{H}}_{\text{el}}(\boldsymbol{\kappa})$ is defined as

\begin{equation}\label{eqn:active_H}
    \hat{\text{H}}_{\text{el}}^{\text{A}}(\boldsymbol{\kappa}) = \sum_{uv}\text{F}^{\text{I}}_{uv}(\boldsymbol{\kappa})\hat{\text{E}}_{uv}+\sum_{uvxy}g_{uvxy}(\boldsymbol{\kappa})\hat e_{uvxy}~,
\end{equation}
where the term $\text{F}^{\text{I}}_{uv}(\boldsymbol{\kappa})$ in \eq{eqn:active_H} is the inactive Fock matrix,

\begin{equation}
\label{eqn:inactive_H}
    \text{F}^{\text{I}}_{uv}(\boldsymbol{\kappa}) = h_{uv}(\boldsymbol{\kappa}) + \sum_i(2g_{iiuv}(\boldsymbol{\kappa})-g_{ivui}(\boldsymbol{\kappa}))~.
\end{equation}
The presence of $\text{F}^{\text{I}}_{uv}(\boldsymbol{\kappa})$ allows to fold into $\hat{\text{H}}^{\text{A}}_{\text{el}}(\boldsymbol{\kappa})$ the mean-field interaction between the active and the inactive electrons.
Following the classification of the electrons, the electronic energy $\text{E}_{\text{el}}$ is also given by the sum of an active, $\text{E}_{\text{el}}^{\text{A}}$, and an inactive, $\text{E}_{\text{el}}^{\text{I}}$, part:

\begin{equation}\label{eqn:energy_split}
    \text{E}_{\text{el}} = \text{E}_{\text{el}}^{\text{I}} + \text{E}_{\text{el}}^{\text{A}}~.
\end{equation}
The $\text{E}_{\text{el}}^{\text{I}}$ component is obtained classically according to

\begin{equation}\label{eqn:energy_inactive}
    \text{E}_{\text{el}}^{\text{I}} = \sum_i h_{ii}(\boldsymbol{\kappa}) + \text{F}^{\text{I}}_{ii}(\boldsymbol{\kappa})~,
\end{equation}
while the $\text{E}_{\text{el}}^{\text{A}}$ component is obtained by measuring the expectation value of $\hat{\text{H}}_{\text{el}}^{\text{A}}$ on the quantum computer.

\subsection{Ansatz optimization and measurement on quantum hardware}\label{sec:oo-VQE}
We define the exact reference for the measurements on the quantum computer by optimizing the ansatz $\ket{\Psi(\boldsymbol{\theta})}$ at the statevector level using the \ac{oo-vqe} algorithm \cite{Mizukami_2020,Sokolov_2020,OO_ADAPT_VQE_2024}. Within \ac{oo-vqe}, the energy minimization problem is formulated in terms of both $\boldsymbol{\theta}$ and $\boldsymbol{\kappa}$, i.e.,

\begin{equation}\label{eqn:oo-vqe-1}
\text{E}_{\text{el}} = \min_{\boldsymbol{\theta}, \boldsymbol{\kappa}} \bra{\Psi(\boldsymbol{\theta})}\hat{\text{H}}_{\text{el}}(\boldsymbol{\kappa})\ket{\Psi(\boldsymbol{\theta})}.
\end{equation}
The optimization of the orbitals modifies the molecular integrals within $\hat{\text{H}}(\boldsymbol{\kappa})$ as

\begin{equation}\label{eqn:one_elect_int_kappa}
	h_{p'q'}(\boldsymbol\kappa) = \sum_{pq} [\exp({\boldsymbol\kappa})]_{p'p}h_{pq} [\exp(-{\boldsymbol\kappa})]_{qq'},
\end{equation}

\begin{equation}\label{eqn:two_elect_int_kappa}
    g_{p'q'r's'}(\boldsymbol\kappa) = \sum_{pqrs} [\exp({\boldsymbol\kappa})]_{p'p} [\exp({\boldsymbol\kappa})]_{q'q} g_{pqrs} [\exp(-{\boldsymbol\kappa})]_{rr'}  [\exp({\boldsymbol\kappa})]_{ss'}.
\end{equation}
The matrix $\boldsymbol\kappa$ contains the parameters of the $\hat{{\kappa}}$ operator which has been used to obtain $\hat{\textrm{H}}_\textrm{el}(\boldsymbol\kappa)$. The $\hat{{\kappa}}$ operator corresponds to the generator of orbital rotations and is defined as

\begin{equation}
    \hat{{\kappa}} = \sum_{pq}{\kappa}_{pq}\Big(\hat{\text{E}}_{pq} - \hat{\text{E}}_{qp}\Big),
\end{equation}
with $\hat{\text{E}}_{pq}\in\{\hat{\text{E}}_{ui},\hat{\text{E}}_{ai},\hat{\text{E}}_{au}\}$. From the \ac{oo-vqe} optimization we obtain the set of optimized ansatz parameters, $\boldsymbol{\theta}^*$, and orbital rotation parameters, $\boldsymbol{\kappa}^*$, corresponding to the minimized electronic energy $\text{E}_{\text{el}}^*$. The measurement on the quantum computer aims at reproducing $\text{E}_{\text{el}}^*$ by estimating the expectation value

\begin{equation}\label{eqn:QH_expectation_value}
    \bra{\Psi(\boldsymbol{\theta}^*)}\hat{\text{H}}_{\text{el}}(\boldsymbol{\kappa}^*)\ket{(\boldsymbol{\theta}^*)} = \text{E}_{\text{el}}^{\text{I}}(\boldsymbol{\kappa}^*) + \bra{\Psi(\boldsymbol{\theta}^*)}\hat{\text{H}}_{\text{el}}^{\text{A}}(\boldsymbol{\kappa}^*)\ket{\Psi(\boldsymbol{\theta}^*)},
\end{equation}
where, as previously mentioned, solely the expectation value of $\hat{\text{H}}_{\text{el}}^{\text{A}}$ is measured on the quantum computer, while $\text{E}_{\text{el}}^{\text{I}}(\boldsymbol{\kappa}^*)$ is obtained classically.

\subsection{The \ac{pp-tUPS} ansatz}\label{sec:tUPS-ansatz}
\begin{figure}[H]
    \centering
    \includegraphics[width=0.7\linewidth]{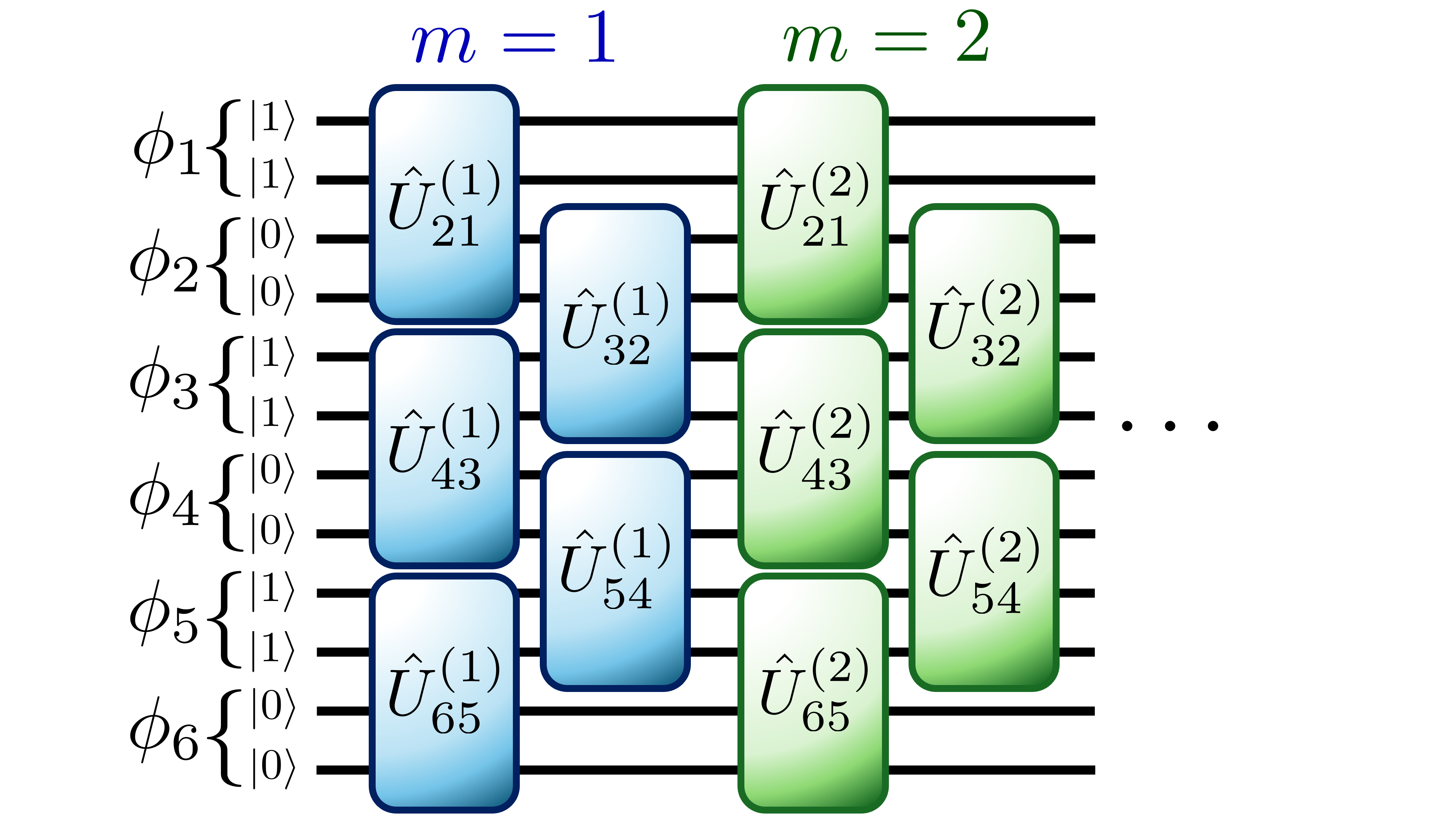}
    \caption{Quantum circuit relative to the pp-tUPS ansatz. A 2-layer example is shown, the first layer corresponding to the blue tiles, the second layer to the green tiles. The curly brackets indicate the qubits associated to a given spatial orbital $\phi_p$. The initial state of the qubit register reflects the grouping of the spatial orbitals associated with the perfect-pairing approximation.}
    \label{fig:pptUPS}
\end{figure}

We adopt the \ac{pp-tUPS} wave function ansatz introduced by Burton \cite{Burton_2024}. This ansatz is defined in terms of the unitary operator

\begin{equation}\label{eqn:tUPS_unitary}
\hat{\text{U}}_{\mathrm{tUPS}}^{(\text{L})}(\boldsymbol{\theta}) = \prod^{\text{L}}_{m=1}\left(\prod^{\text{R}}_{j=1}\hat{\text{U}}^{(m)}_{2j+1,2j}\prod^{\text{S}}_{j=1}\hat{\text{U}}^{(m)}_{2j,2j-1}\right)~,
\end{equation}
where, given N spatial orbitals, $\text{R}=\frac{\text{N}-2}{2}$ and $\text{S}=\frac{\text{N}}{2}$ for N even, or $\text{R}=\frac{\text{N}-1}{2}$ and $\text{S}=\frac{\text{N}-1}{2}$ for N odd. For each layer $m$, the operator $\hat{\text{U}}^{(m)}_{pq}$ is defined in terms of one- and two-body operators as

\begin{equation}\label{eqn:tUPS_tiles_definition}
\hat{\text{U}}^{(m)}_{pq} 
= \exp\left(\theta^{(m)}_{pq,1} \hat{\kappa}^{(1)}_{pq} \right) \exp\left(\theta^{(m)}_{pq,2} \hat{\kappa}^{(2)}_{pq} \right) \exp\left(\theta^{(m)}_{pq,3} \hat{\kappa}^{(1)}_{pq} \right),
\end{equation}
where $\hat{\kappa}^{(1)}_{pq} = \hat{\textrm{E}}_{pq}-\hat{\textrm{E}}_{qp}$, $\hat{\kappa}^{(2)}_{pq} = \hat{\text{E}}^2_{pq}-\hat{\text{E}}^2_{qp}$, and $\{\theta^{(m)}_{pq,1}, \ \theta^{(m)}_{pq,2}, \ \theta^{(m)}_{pq,3}\}$ is the set of variational parameters. As illustrated in \figref{fig:pptUPS}, each unitary $\hat{\text{U}}^{(m)}_{pq}$ defines a `tile' in the quantum circuit, with each new layer being composed by two-tiled column units. The construction of the \ac{pp-tUPS} ansatz starts from the initialization of the qubit registers in terms of alternating couples of occupied and empty spin orbitals (which correspond, in turn, to the alternation of occupied and empty spatial orbitals according to the perfect pairing approximation \cite{Hurley_1953}). 
A circuit layer $m$ is defined by intercalating $\hat{\text{U}}_{pq}^{(m)}$ tiles---each acting on four spin orbitals---to fill the entire qubit space. The layer structure is repeated L times systematically improving the ansatz and converging, for $\text{L}\to\infty$, to the exact limit \cite{Burton_2024}. Therefore, we express the wave function ansatz as

\begin{equation}
\ket{\Psi(\boldsymbol{\theta)}} = \hat{\text{U}}^{(\text{L})}_{\mathrm{tUPS}}(\boldsymbol{\theta}) \ket{0011\cdots 0011},
\end{equation}
where we regulate the number of tUPS layers based on the desired level of accuracy and the capabilities of the quantum hardware.

\subsection{Error Mitigation Using QESEM}\label{sec:QESEM-em}

Quantum Error Suppression and Error Mitigation (QESEM) \cite{aharonov2025reliable} is a characterization-based framework for obtaining reliable, high-accuracy results from noisy quantum circuits.
Before mitigating a given circuit, QESEM first measures the actual noise affecting the device and uses this information to guide every later step, rather than assuming a fixed, generic noise model. Built on an unbiased quasi-probabilistic (QP) mitigation framework, QESEM preserves the reliability of QP-based mitigation methods while introducing key innovations that improve efficiency and scalability. This is particularly relevant for molecular VQE applications, where high accuracy is required and the target observable may involve many Pauli terms that span across multiple measurement bases. Below is a short summary of the method (see Ref.~\cite{aharonov2025reliable} for extensive details).

\subsubsection{Workflow Overview}

The QESEM workflow consists of five interconnected stages:

\noindent\textbf{Device Characterization:} A preliminary characterization of the whole device identifies suppressible coherent errors and maps the fidelities of the available device operations. 
This step provides the real-time calibration data used by the noise-aware transpilation and error suppression stages.

\noindent\textbf{Noise-Aware Transpilation:} Multiple transpilations of the input circuit are constructed, and the one with the lowest estimated characterization and mitigation \ac{QPU} time is selected. 

\noindent\textbf{Error Suppression:} Native device operations are reconfigured to improve fidelity and predictability in error-mitigated execution, for example by reversing coherent errors and applying Pauli twirling. Pauli twirling converts coherent errors and off-diagonal dissipative errors into simpler stochastic Pauli channel noise, which is easier to characterize and mitigate reliably in the next stages.

\noindent\textbf{Circuit Characterization:} The errors affecting the two-qubit gate layers in the chosen transpilation are characterized and used to construct and fit a tailored local error model, so that the noise acting on the specific circuit at hand -- rather than a generic, device-wide estimate -- enters the mitigation step.

\noindent\textbf{Error Mitigation:} QP decompositions are constructed from this error model and sampled in an adaptive drift-robust flow to obtain an estimate and error bar for each requested observable. This final stage is typically the most resource-intensive, since obtaining a small statistical error bar requires sampling many circuit variants.

\subsubsection{Key Aspects}

The key aspects of QESEM are:

\noindent\textbf{Multi-Type QP Decompositions:} Efficient multi-type quasi-probabilistic decompositions mitigate errors in both Clifford and fractional (non-Clifford) two-qubit gates. This approach can reduce circuit depths by up to $2\times$ compared to methods that compile fractional gates into Clifford gates, since it avoids the extra Clifford-gate overhead that such a compilation would otherwise introduce.

\noindent\textbf{Active Volume Identification:} QESEM employs a commutativity lightcone algorithm to identify which gates affect the measured observable and focus mitigation efforts only on those gates. Intuitively, not every gate in a circuit influences a given observable's expectation value; errors on gates outside this lightcone do not need to be mitigated. This helps reduce \ac{QPU} runtime overhead by avoiding mitigation of gates outside the active volume of the observable.

\noindent\textbf{Resource estimation and minimization:} Empirical variance estimation is used to determine the resources required to reach a requested statistical accuracy and to optimize resource allocation throughout the execution, so that QPU time is not spent beyond what is needed for the requested precision.

\noindent\textbf{Measurement-basis optimization:} For target observables containing non-commuting terms -- as is generally the case for a molecular Hamiltonian expressed as a sum of Pauli operators -- QESEM automatically finds a set of local measurement bases required to measure the observable, heuristically trying to minimize the number of bases. QESEM allocates resources across bases by estimating the variance of the measured observables in each basis and then choosing the optimal precision per basis in order to minimize the required \ac{QPU} time such that the statistical error of the total target observable is less than the requested precision.

\noindent\textbf{Drift Mitigation:} Device noise can change during an experiment. Characterization and mitigation circuits are interleaved so that the noise model, QP distribution, and retroactive corrections can be updated during execution, keeping the mitigation consistent with the device's noise as it evolves over the course of a run.

\noindent\textbf{Parallelization:} For narrow circuits utilizing only a small subset of the device qubits, automatic patch parallelization can map the circuit to different qubit subsets and execute them simultaneously, reducing QPU overhead. In this article, for most bond lengths, this capability was leveraged by distributing the workload across 4 QPU regions rather than mitigating a single patch.

In \ac{qem} methods in general, and in \ac{qesem} in particular, bias in observables expectation values due to hardware noise is eliminated at the cost of increased variance of the mitigated expectation value estimator, which leads to a QPU runtime overhead for achieving a given statistical error compared to raw noisy execution. 
In \ac{qesem} the user sets the required statistical error target for the mitigated observable by the \textbf{precision} parameter, which we will denote here as $P$. In the case herein, this precision will specify the uncertainty range  for the mitigated molecular ground state energy. The mitigation procedure will stop once the desired precision is achieved. The stricter the precision (small values of $P$), the larger the required \ac{QPU} runtime, as it scales as the inverse squared of the precision, $T_\text{QPU} \propto \tfrac{1}{P^2}$.


\section{Computational Methods \label{comp_meth}}

\subsection{Computational protocol}
The computational procedure aims at reproducing the \ac{PES} corresponding to the symmetric stretching mode (i.e. the O-H bonds are shortened/stretched in equal measure) of the water molecule, bent in a 104.5 degrees angle. At each geometry, we obtained $h_{\text{nuc}}$, summing it to the corresponding value of $\text{E}_{\text{el}}^*$ estimated on the quantum computer. For all calculations of $\text{E}_{\text{el}}^*$ we considered a (4,4) active space (i.e., 4 electrons in 4 orbitals) and the STO-3G basis. We obtained the Hamiltonian $\hat{\text{H}}_{\text{el}}^{\text{A}}$ using the implementation by \citeauthor{Rossmannek_2021} \cite{Rossmannek_2021} included in \textsc{Qiskit nature} 0.7.2 \cite{qiskit2024}. At each geometry, we obtained the initial electronic integrals composing $\hat{\text{H}}_{\text{el}}^{\text{A}}(\boldsymbol{\kappa})$ through the \textsc{PySCF} quantum chemistry package \cite{sun2007python,sun2018pyscf,pySCF}. We first optimized the 1-layer \ac{pp-tUPS} ansatz and the orbital parameters at the statevector level through the \ac{oo-vqe} algorithm; we performed this task using the \textsc{SlowQuant} package \cite{SlowQuant}, while adopting the \ac{bfgs} algorithm from \textsc{scipy} \cite{2020SciPy-NMeth}. 

As a second step, we mapped $\hat{\text{H}}_{\text{el}}^{\text{A}}(\boldsymbol{\kappa}^*)$ and the optimized 1-layer pp-tUPS ansatz to the qubit space via the Jordan-Wigner transformation \cite{Jordan_Wigner_1928} as implemented in \textsc{Qiskit nature} 0.7.2 \cite{qiskit2024}. At each considered geometry, we estimated the active space energy $\text{E}_{\text{el}}^\textrm{A}$ via the \textsc{Qesem} error-mitigation API \cite{aharonov2025reliable} on IBM's Aachen quantum computer. The specifics on the Aachen hardware are enlisted in \tabref{tab:ibm_aachen}.
\begin{table}[H] 
	\centering
	\caption{IBM's Aachen hardware overall performances. \label{tab:ibm_aachen}}
	\begin{tabular}{cccccc}
		\toprule
		\textbf{\makecell{Number of\\ qubits}} & \textbf{Couplers} & \textbf{\makecell{Median 2 Qubit\\ error}} & \textbf{\makecell{Mean 2 Qubit error\\ (layered)}} & \textbf{\makecell{2 Qubit error for\\ 8 qubits (layered)} } &  \textbf{\makecell{Processor\\ type}} \\ \addlinespace[0.3em]
		156 & 176 & $1.46\times10^{-3}$ & $ 2.97\times 10^{-3}$ & $1.68 \times 10^{-3}$ &  Heron r3 \\ 
		\bottomrule
	\end{tabular}
	\tabletext{Note: Values taken on the 15$^\text{th}$ April 2026. These values are sensible to calibration and may vary. }
\end{table}
We obtain an estimate for $\text{E}_{\text{el}}^*$ at each geometry by summing to $\text{E}_{\text{el}}^{\text{A}*}$ the inactive contribution $\text{E}_{\text{el}}^{\text{I}*}$ obtained at the end of the \ac{oo-vqe} simulation on the classical simulator. 

\subsection{Merging multiple same geometry results}
When using the QESEM method, a single run yields a mitigated energy $\text{E}_i$ with its associated \ac{std} $\sigma_i$. Due to the statistical error of the mitigated energy, different independent outcomes can arise from different runs of otherwise identical calculations. Merging the results from different runs has the advantage of accumulating the shot information, leading to better statistical representation and thus a more meaningful final result. For this purpose, we employ the inverse-variance weighted average according to which

\begin{equation} \label{eq:E_merged}
    \text{E}_\text{merged}=\frac {\sum _{i} \text{E}_{i}/\sigma_{i}^{2}}{\sum_{i}1/\sigma_{i}^{2}} \, ,
\end{equation}
with the associated \ac{std}
\begin{equation} \label{eq:sigma_merged}
    \sigma^2(\text{E}_\text{merged})=\frac {1}{\sum_{i}1/\sigma_{i}^{2}}\,.
\end{equation}
In the following, for the same type of calculations, different run batches are performed, with the $n^\text{th}$ batch being labeled as R$n$.

\subsection{Definition of the wave-function ansatz}

We adopt a one-layer \ac{pp-tUPS} ansatz in consideration of its sufficiently accurate reproduction of \ac{PES} (see results section). At the pp-tUPS(1)/(4,4)/STO-3G level of theory, the ansatz operator reads 

\begin{equation}\label{eqn:tUPS_ansatz_experiment}
\begin{aligned}
	\hat{\text{U}}_\text{pp-tUPS}^{(1)}(\boldsymbol \theta) &= \hat{\text{U}}^{(1)}_{32} \hat{\text{U}}^{(1)}_{43} \hat{\text{U}}^{(1)}_{21} \\
    &= \underbrace{\exp\left(\theta^{(m)}_{32,1} \hat{\theta}^{(1)}_{32} \right)}_\text{dropped}  \exp\left(\theta^{(m)}_{32,2} \hat{\theta}^{(2)}_{32} \right) \exp\left(\theta^{(m)}_{32,3} \hat{\theta}^{(1)}_{32} \right) \hat{\text{U}}^{(1)}_{43} \hat{\text{U}}^{(1)}_{21}.
\end{aligned}
\end{equation}
Since the orbital optimization already accounts for the singlet single excitation within the active space, the corresponding tile of the tUPS ansatz is redundant within the oo-VQE framework. We define the ansatz $\ket{\Psi(\boldsymbol{\theta})}$ according to \eq{eqn:tUPS_ansatz_experiment}, while initializing the qubit register to the perfect pairing ordering. The characteristics of the circuit are summarized in \tabref{tab:pptUPS_ansatz_details}. A graphical representation of the circuit is displayed in \figref{fig:pptUPS(1)_AS(4,4)} of the \ac{SI}.
\begin{table}[H] 
	\centering
	\caption{Details on the circuit for the water molecule using the pp-tUPS(1)/(4,4)/STO-3G level of theory. The entanglers are the CNOT and the CZ gates for the ideal and the transpiled circuits, respectively. \label{tab:pptUPS_ansatz_details}}
	\begin{tabular}{lcc}
		\toprule
		 & \textbf{Ideal circuit} &  \textbf{Transpiled circuit} \\
		 Number of qubits & 8  & 8 \\ 
		 Entanglers  & 62  &  116 \\ 
	   Depth & 64  &  114\\
		\bottomrule
	\end{tabular}
\end{table}


\subsection{Absolute relative error}

As mentioned in Sec. \ref{sec:oo-VQE}, the measurements on quantum hardware involve solely the expectation value of $\hat{\text{H}}_{\text{el}}^{\text{A}}$, which corresponds to the active space energy $\text{E}_{\text{el}}^{\text{A}}$. For the measurements on quantum hardware, $\hat{\text{H}}_{\text{el}}^{\text{A}}$ is mapped onto the qubit basis assuming the form of a weighted sum of Pauli strings. The expectation value measurement on the quantum computer focuses on the \textit{traceless} part of $\hat{\text{H}}_{\text{el}}^{\text{A}}$, which excludes the contribution of the identity Pauli string and yields the traceless energy $\text{E}_{\text{el, traceless}}^{\text{A}}$. The value of $\text{E}_{\text{el}}^{\text{A}}$ is retrieved in a classical post-processing step by summing to $\text{E}_{\text{el, traceless}}^{\text{A}}$ the contribution of $\hat{\text{H}}_{\text{el}}^{\text{A}}$'s term associated with the identity Pauli string.

The \ac{qesem} mitigation focuses on achieving a certain precision $P$, that after calculation results in a \ac{std} $\sigma$ for the mitigated energy. Therefore, as an indicator of the quality of the mitigation, we adopt the absolute relative error

\begin{equation}
    \delta = \frac{\sigma}{\abs{\text{E}_{\text{el, traceless}}^{\text{A}}}} \, ,
\end{equation}
which witnesses the relative difficulty to achieve the desired precision given the magnitude of the retrieved energy.

\section{Results and Discussion \label{results_disc}}

As a first step, we present preliminary classical simulations defining the ideal reference \ac{PES} at the considered level of theory. As a second step, we present the results relative to \ac{PES}s measured on real quantum hardware at two different levels of precision, which make use of the QESEM method as a \ac{qem} backend. The performance and results are analyzed within each subsection. 

\subsection{Preliminary Simulations}

In this section, we compare the \ac{PES} calculated via statevector simulations adopting the one-layer \ac{pp-tUPS} ansatz with that obtained from classical CASSCF simulations (performed with \textsc{PySCF}). Each point of the \ac{PES}s corresponds to the expectation value E of $\hat{\text{H}}$ (cf. Eq. \ref{eqn:electronic_H}) for each nuclear configuration. As discussed in Sec. \ref{sec:H_active_space}, for the \ac{pp-tUPS} \ac{PES} each value
is 

\begin{equation}\label{eqn:energy_split_nuc}
    \text{E} = \text{E}_{\text{el}}^{\text{I}} + \text{E}_{\text{el}}^{\text{A}} + \text{E}_{\text{nuc}}.
\end{equation}
Via the statevector simulations we determine $\text{E}_{\text{el}}^{\text{I}}$ and $\text{E}_{\text{nuc}}$ at each point of the \ac{PES}. In the estimation of E on quantum hardware, $\text{E}_{\text{el}}^{\text{I}}$ and $\text{E}_{\text{nuc}}$ thus represent constants which are added to the corresponding measurements of $\text{E}_{\text{el}}^{\text{A}}$.

\begin{figure}[h]
    \centering
    \includegraphics[width=0.7\linewidth]{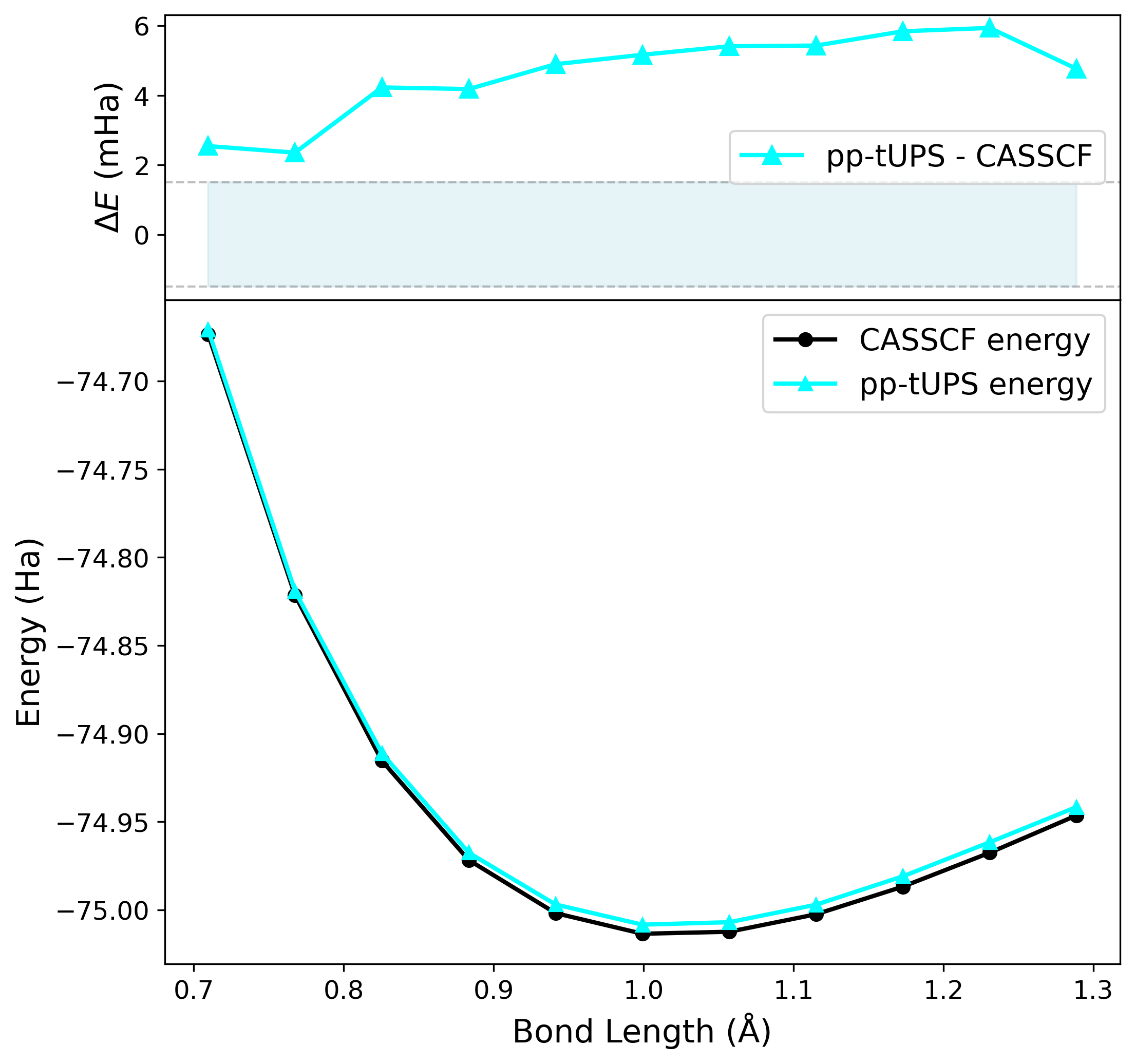}
    \caption{Comparison of CASSCF and one-layer pp-tUPS PESs relative to the symmetric stretching mode of water. (Bottom) Total energy from CASSCF (black-circle) and one layer pp-tUPS (cyan-triangle) ideal simulation. (Top) pp-tUPS(1) energy error with respect to the CASSCF energy. The blue area represents the chemical accuracy zone with respect to the CASSCF reference.}
    \label{fig:CAS_vs_pptUPS_error}
\end{figure}

The comparison of the calculated \ac{PES}s is shown in \figref{fig:CAS_vs_pptUPS_error}.
The ideal one-layer \ac{pp-tUPS} energies closely follow the CASSCF results, overestimating the energy of the latter by around $5$ mHa. The error is close to the ``chemical accuracy'' region of $1.5$ mHa and it is considerably smaller compared to usual error for this system size $(\sim 500$ mHa, \textit{vide infra}). While additional \ac{pp-tUPS} layers bring the statevector \ac{PES} ever closer to the CASSCF reference, they require progressively deeper circuits. Since the goal of this study is to evaluate \ac{qesem}'s capabilities in enabling the reproduction of the statevector reference, we considered the one-layer pp-tUPS ansatz as a reference for the upcoming calculations on real \ac{QPU}. This provides us with a good compromise between circuit depth and accuracy, here defined as closeness to the classical CASSCF reference.

\clearpage

\subsection{Quantum Experiment}

In this section, we present the \ac{PES}s estimates on the quantum hardware for two different precision parameters. For both cases, the \ac{PES} is retrieved and analyzed, while also quantifying the required computational resources.

\subsubsection{Loose precision}

\begin{figure}[H]
    \centering
    \includegraphics[width=0.7\linewidth]{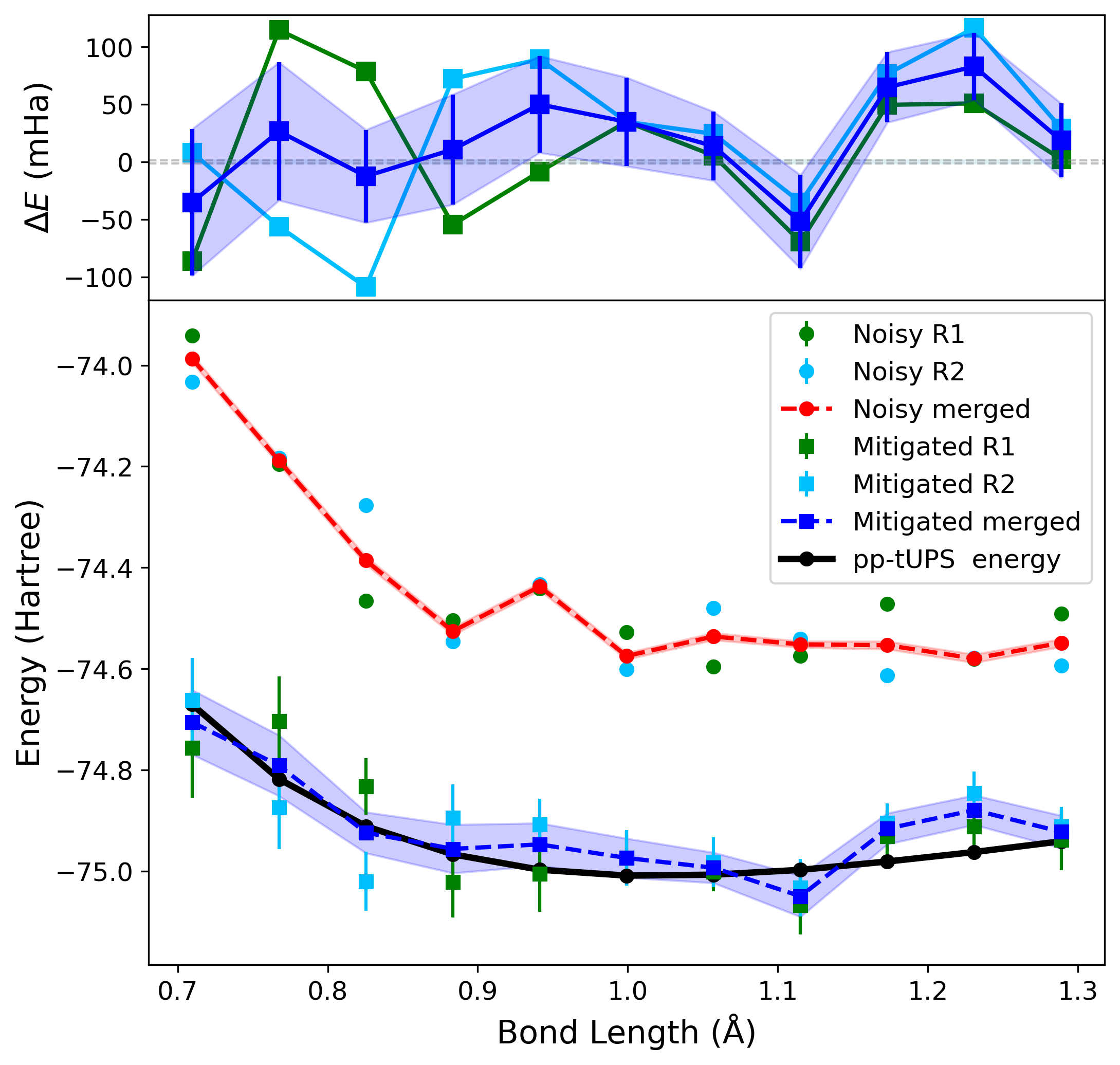}
    \caption{Water \ac{PES} at the pp-tUPS(1)/(4,4)/STO-3G level of theory for a 0.1 precision. (Bottom) The noisy and mitigated energies are plotted with circles and squares, respectively. Their standard deviations are drawn as vertical lines. The merged results are the dotted lines embedded within their standard deviation colored zone. The pp-tUPS reference is in black. (Top) The error graph of the mitigated energies with respect to the pp-tUPS ideal results. The barely visible light blue region is the ``chemical accuracy'' zone.}
    \label{fig:Averaged_Error_PES_P01}
\end{figure}

For the first two run batches, R1 and R2, the precision parameter is set to $0.1$ (Ha). For each considered geometry, two independent full measurements with the QESEM program are performed. Moreover, two different PES are associated to each run: the noisy PES corresponds to the raw energies measured on the quantum computer without QEM; the mitigated PES is formed by QESEM's mitigated energies. The results are shown in \figref{fig:Averaged_Error_PES_P01}.
For both R1 and R2, the noisy results consistently overestimate the energy by about $500$ mHa. The noisy energies exhibit no smooth surface shape and no discernible pattern across runs. Turning to the mitigated results, the energies lie considerably closer to the pp-tUPS reference, though they intermittently under- or overestimate the target energy. The maximum errors are on the order of 100 mHa, with all remaining points staying within this range. The R1 and R2 values are combined using Eqs.~\eqref{eq:E_merged} and~\eqref{eq:sigma_merged} yielding the merged noisy (red) and merged mitigated (dark blue) results. The mitigated weighted average is taken as the main representative result; it inherits the irregular over- and underestimation behavior, with occasional compensation when R1 and R2 deviate in opposite directions. Across the considered eleven geometries, only 3 reference energies lie outside the merged uncertainty domain (blue zone). 

\begin{figure}[H]
    \centering
    \includegraphics[width=0.8\linewidth]{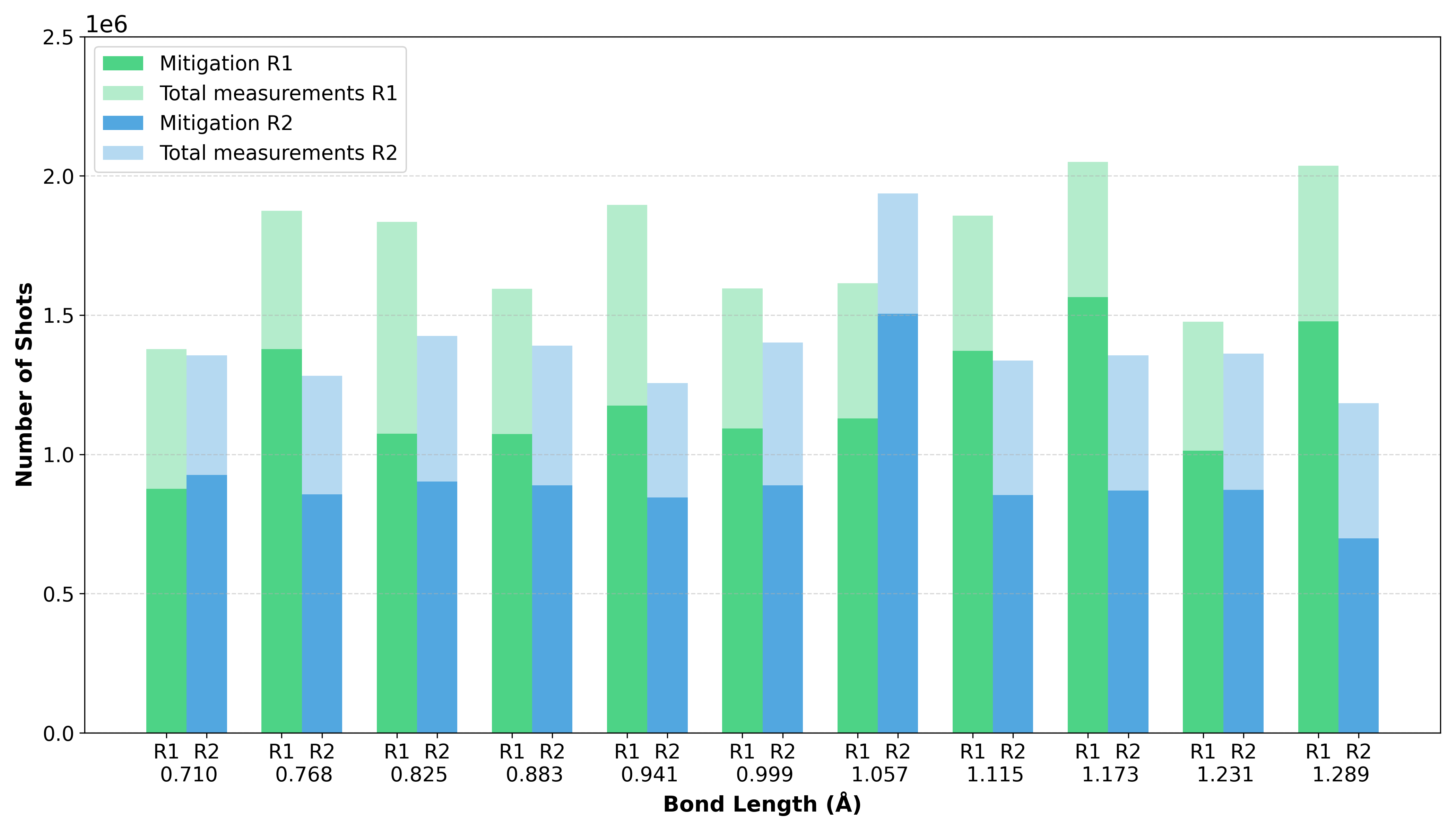}
    \caption{Number of shots for R1 and R2, and for each geometry. The mitigation shots (darker colors) characterize the noise whereas the Total measurements(lighter colors) include device familiarization and calibration.}
    \label{fig:Shots_P01}
\end{figure}

As previously explained, due to hardware instabilities and intrinsic statistical behavior, achieving the required precision of $0.1$ hartrees may require different number of shots from one run to another. The number of shots for each run and geometry, for the mitigation scheme and the actual observable measurement are depicted in \figref{fig:Shots_P01}.
On average, the QESEM algorithm achieves $0.1$ precision using about $1.6$ million shots in total. As expected, the number of shots varies even independently of the run or geometry. Each run is unique and adapts to the estimated noise model of the moment. The most extreme example of this can be seen for the last geometry ($1.289$ \AA), where the same calculation took about $1\,200\,000$ shots for R2, and as much as $2\,040\,000$ shots for R1. 
In terms of \ac{QPU} time, it corresponds to $8$ minutes versus $11$ minutes of utilizing the quantum hardware, see \figref{fig:QPUTime_P01} in \ac{SI}. 

A particular strength of \ac{qesem}'s approach is its adaptability, which conveniently allows to mitigate the errors depending on their magnitude and particular behavior. While the merged mitigated results never fall outside a 100 mHa interval from the classical reference, a precision of 0.1 Ha is usually insufficient for chemical applications. Thus, our next step was to repeat the experiments with a more strict precision.   

\clearpage

\subsubsection{Tight precision}

\begin{figure}[H]
    \centering
    \includegraphics[width=0.7\linewidth]{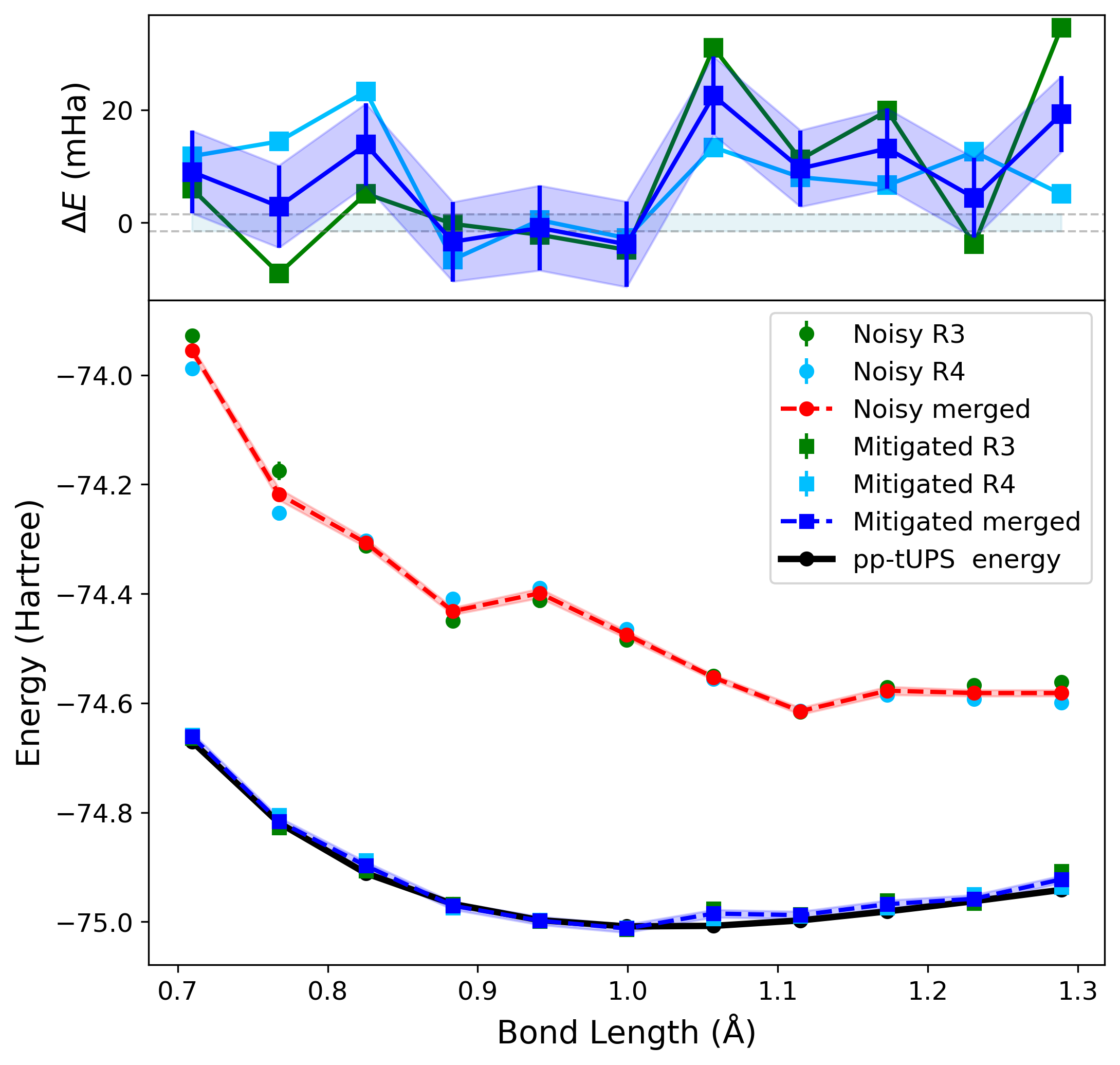}
    \caption{Water \ac{PES} at the pp-tUPS(1)/(4,4)/STO-3G level of theory for a 0.01 precision. (Bottom) The noisy and mitigated energies are plotted with circles and squares, respectively. Their standard deviations are drawn as vertical lines. The merged results are the dotted lines embedded within their standard deviation colored zone. The pp-tUPS reference is in black. (Top) The error graph of the mitigated energies with respect to the pp-tUPS ideal results. The light blue region is the ``chemical accuracy'' zone.}
    \label{fig:Averaged_Error_PES_P001}
\end{figure}

Searching for a better mitigation, and expecting to approach the golden standard of chemical accuracy, the estimations are repeated with a more stringent precision.
The next two set of measurements, R3 and R4, are performed using the 0.01 (Ha) precision value. Similarly to the results presented in Fig.~\ref{fig:Shots_P01}, the results of each run (in cyan and green) and the merged result (blue) are depicted in \figref{fig:Averaged_Error_PES_P001}.
The raw (noisy) results behavior are practically identical to those obtained in the loose precision case, with the sole difference that the raw R3 and R4 values are appreciably closer one to another. This is because, on average, for the raw measurement part the R1-R3 experiment used $\sim 500\,000$ shots, whereas the R3-R4 experiment required $\sim 4\,300\,000$ shots. Increasing the number of shots diminishes the shot noise, which in turn provides a more faithful statistical representation of the expected outcome.

The R3 and R4 mitigated results are considerably closer to the ideal energies. The maximum error of $34$ mHa occurs for R3's longest bond length energy, while all other mitigated energies remain within $30$ mHa of the reference. 
This error magnitude is comparable to other recent results \cite{ziems2024options,reinholdt2025critical,Belaloui_2025,rasmussen2026jctc} where similar energy errors are found, although for smaller systems and active spaces. However, for the system and active space considered here, one can point out a significant improvement compared to the $\sim 200$ mHa error of the same system in ref.~\cite{hernandez2026analytical}.
Furthermore, the mitigated results are merged once again, yielding energies 
in-between the original R3 and R4 results. The compensating nature of the merged PES leads to errors within $20$ mHa of the ideal reference. Although the distribution is stochastic, and results herein should be taken as indicative rather than definitive, it is worth noting that one merged energy (at $0.941$ \AA) falls directly within chemical accuracy. If we consider the standard deviation range of $\pm 5$ mHa, corresponding to half the demanded precision, then five out of the eleven merged mitigated energies lie within chemical accuracy.

\begin{figure}[H]
    \centering
    \includegraphics[width=0.8\linewidth]{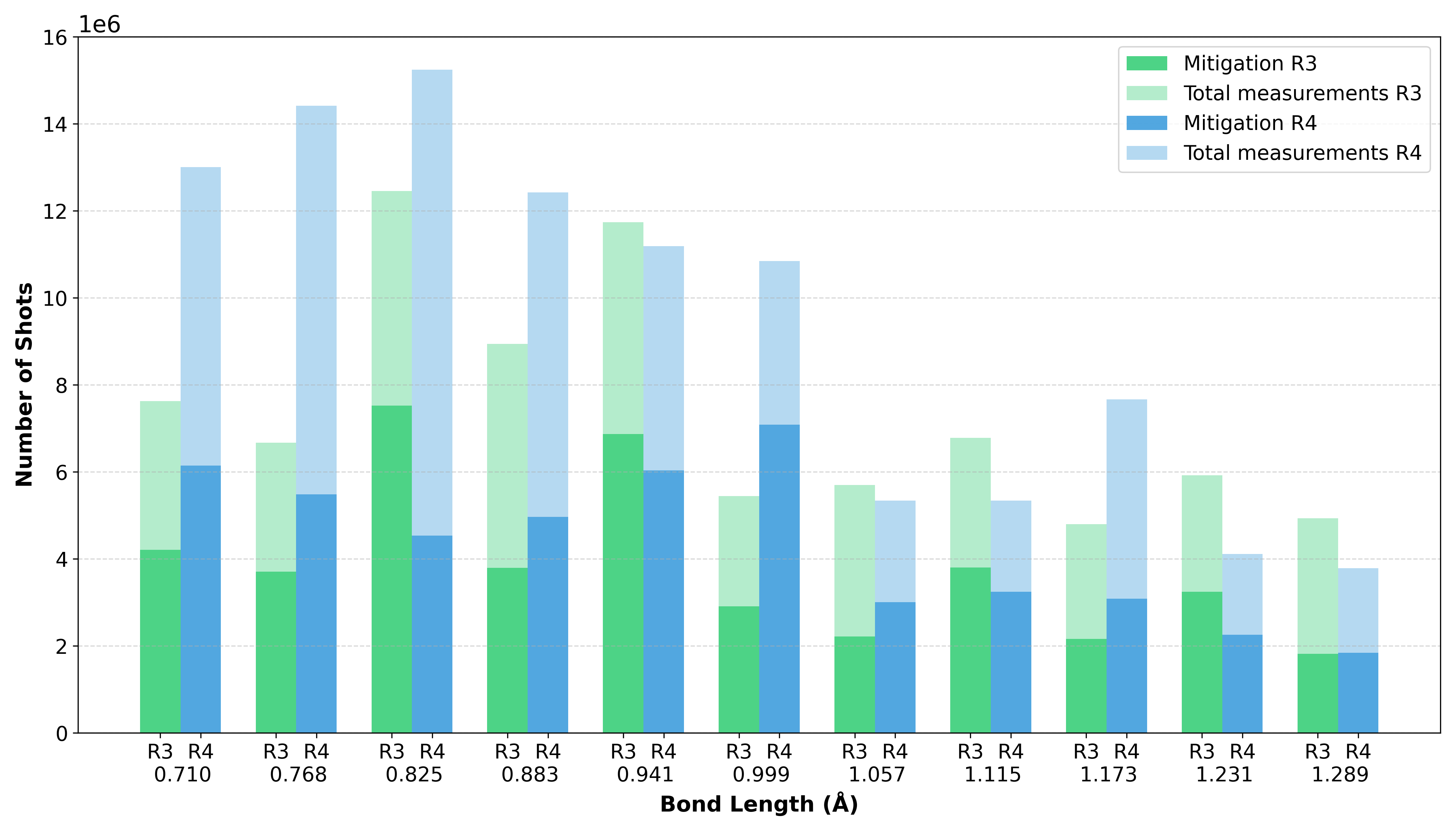}
    \caption{Number of shots for run R3 and R4, and for each geometry. The mitigation shots (darker colors) characterize the noise whereas the Total measurements (lighter colors) include device familiarization and calibration.}
    \label{fig:Shots_P001}
\end{figure}

Reaching this level of precision required, on average, 5 times more shots than the loose precision results, see \tabref{tab:MaxShots_P01_P001}. 
When individually examining the total number of shots for both runs, depicted in \figref{fig:Shots_P001}, it is clear that the number of shots may strongly vary between runs for the same geometry. The clearest example of this is for the second shortest bond geometry ($0.768$ \AA), where the second run required $2.2$ times more shots than the first run. Once again, the necessary computing resources depend on the noise level and error characterization of the moment. The method thus adapts, demanding more or less shots, to ensure achieving the requested standard deviation. In terms of QPU time resources, see \figref{fig:QPUTime_P001} in \ac{SI}, individual results spanned from half an hour, up to two hours and a half. 
\begin{table}[H] 
    \centering
    \caption{Total number of shots for the merged results.  \label{tab:MaxShots_P01_P001}}
    \begin{tabular}{ccc}
    \toprule
    \textbf{Bond Length} & \multicolumn{2}{c}{\textbf{Precision}} \\
          & 0.1         & 0.01 \\
          \midrule
    0.710 & 2\,733\,400 & 20\,632\,814 \\
    0.768 & 3\,157\,400 & 21\,091\,511\\
    0.825 & 3\,260\,264 & 27\,701\,701\\
    0.883 & 2\,984\,400 & 21\,366\,845\\
    0.941 & 3\,152\,400 & 22\,923\,394\\
    0.999 & 2\,998\,400 & 16\,287\,356\\
    1.057 & 3\,550\,400 & 11\,038\,388 \\
    1.115 & 3\,194\,400 & 12\,129\,350\\
    1.173 & 3\,405\,400 & 12\,465\,090\\
    1.231 & 2\,838\,400 & 10\,034\,589\\
    1.289 & 3\,220\,264 & 8\,720\,817\\
    \midrule
    \textbf{Average} & 3\,135\,921  & 16\,762\,896 \\
    \textbf{$\pm$ Std} &  $\pm$ 226\,344 &  $\pm$ 5\,995\,459 \\
    \bottomrule
    \end{tabular}
\end{table}

\begin{figure}[H]
    \centering
    \includegraphics[width=0.8\linewidth]{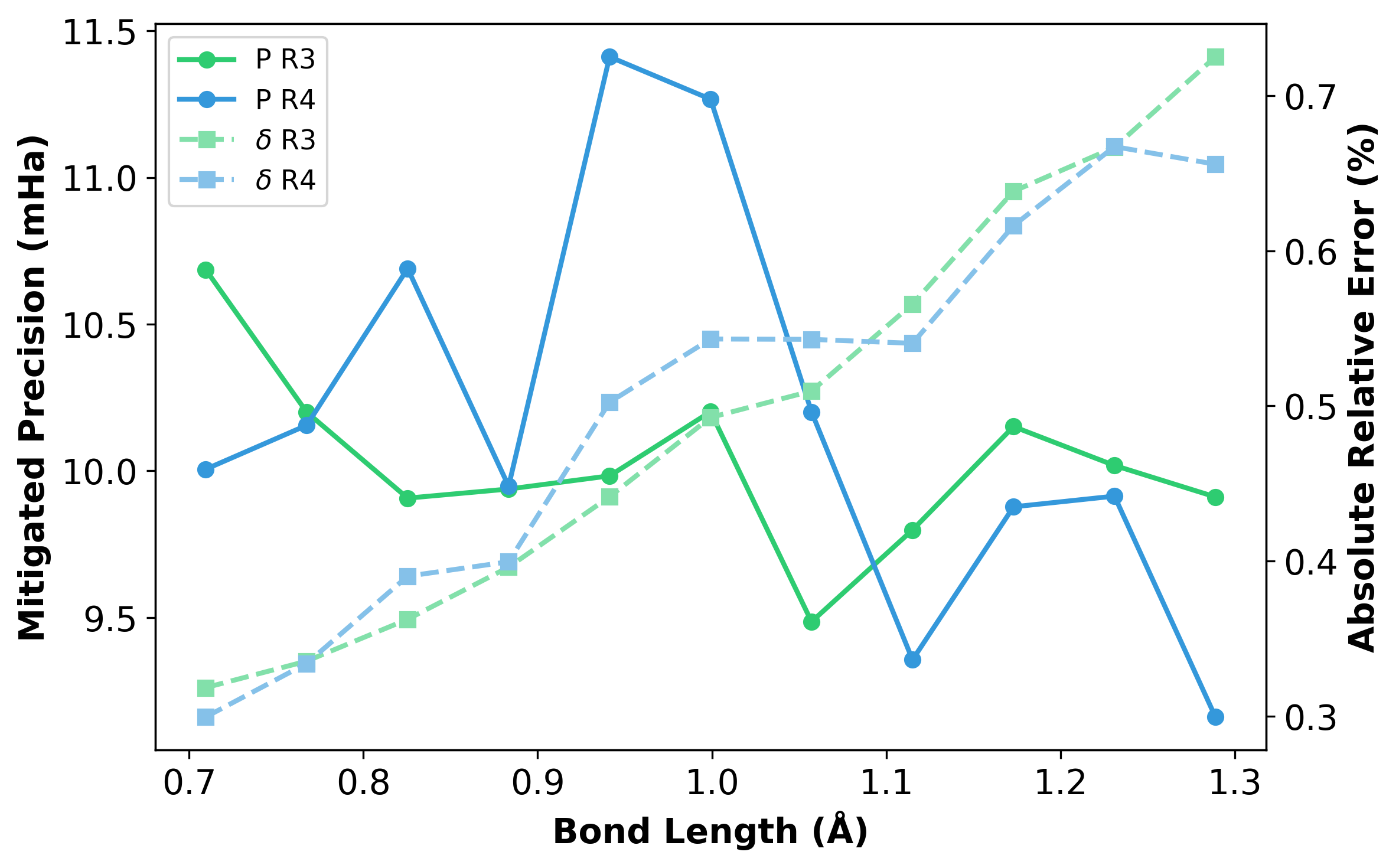}
    \caption{R3 and R4 precisions and absolute relative errors for the different geometries.}
    \label{fig:Precision_Error_R3R4}
\end{figure}

For a deeper analysis, let us now focus in the yielded precisions. The mitigated standard deviation (precision) and the absolute relative error for each point on the PES are depicted in \figref{fig:Precision_Error_R3R4}.
The standard deviations for the mitigated energies respect the $0.01$ (Ha) precision up to the requested significant value. Then, the absolute relative error grows with the bond length. This is due to the fact that the electronic energy diminishes in absolute value for increasing bond lengths, and the precision remains relatively constant.

\section{Conclusions and Outlook \label{conclusions}}

In this work, we presented measurements of the symmetric-stretching PES of water performed on IBM's Aachen quantum hardware using the QESEM mitigation method. We considered a one-layer \ac{pp-tUPS} ansatz defined within a (4,4) active space and the STO-3G basis. We optimized the ansatz at the statevector level via the oo-VQE algorithm, accordingly defining an exact reference for the quantum-hardware estimations. The one-layer \ac{pp-tUPS} ansatz showed a favorable compromise between accurate reproduction of the CASSCF classical reference and circuit depth. We mapped the optimized one-layer \ac{pp-tUPS} ansatz into an 8-qubit register, obtaining a circuit of considerable depth. For both loose (0.1 Ha) and tight (0.01 Ha) precisions two batches of runs were performed, showing both the individual and the merged collective performance of the method. Depending on the demanded precision, the QPU raw measurements typically overestimated the energy by $500$ mHa, whereas the mitigated energies fell under $\sim100$ mHa, under $\sim30$ mHa, or even within chemical accuracy from the statevector reference. Overall, the results improved systematically as the precision parameter was tightened, achieving results comparable to or better than other values reported in the literature \cite{jones2024ground,hernandez2026analytical,rasmussen2026jctc}. For every emulated run, and its associated merged point, the number of shots was shown, quantifying the necessary computing resources. While the results show \ac{qesem}'s potential of achieving highly accurate results on \ac{NISQ} quantum hardware, they also highlight the significant shots overhead required to achieve the expected precision.

\section{Acknowledgments}
We acknowledge financial support from Innovation Fund Denmark 
(Project nos. 4340-00005B and 4340-00006B) 
and Israel Innovation Authority for the Eureka Project ``Q-Chemion''. 
S.P.A.S., S.C., J. K., K.M.Z., and E. K. also acknowledge financial support from the Novo Nordisk Foundation (NNF) for the focused research project “Hybrid Quantum Chemistry on Hybrid Quantum Computers” (Grant no. NNFSA220080996). The authors also acknowledge the scientific contributions of Ori Alberton and Netanel H. Lindner from Qedma.


\clearpage
\section{Supporting Information}

\renewcommand*{\thefigure}{S\arabic{figure}}
\setcounter{figure}{0}  
\setcounter{table}{0}

\begin{figure}[h]
    \centering
    \includegraphics[width=0.9\linewidth]{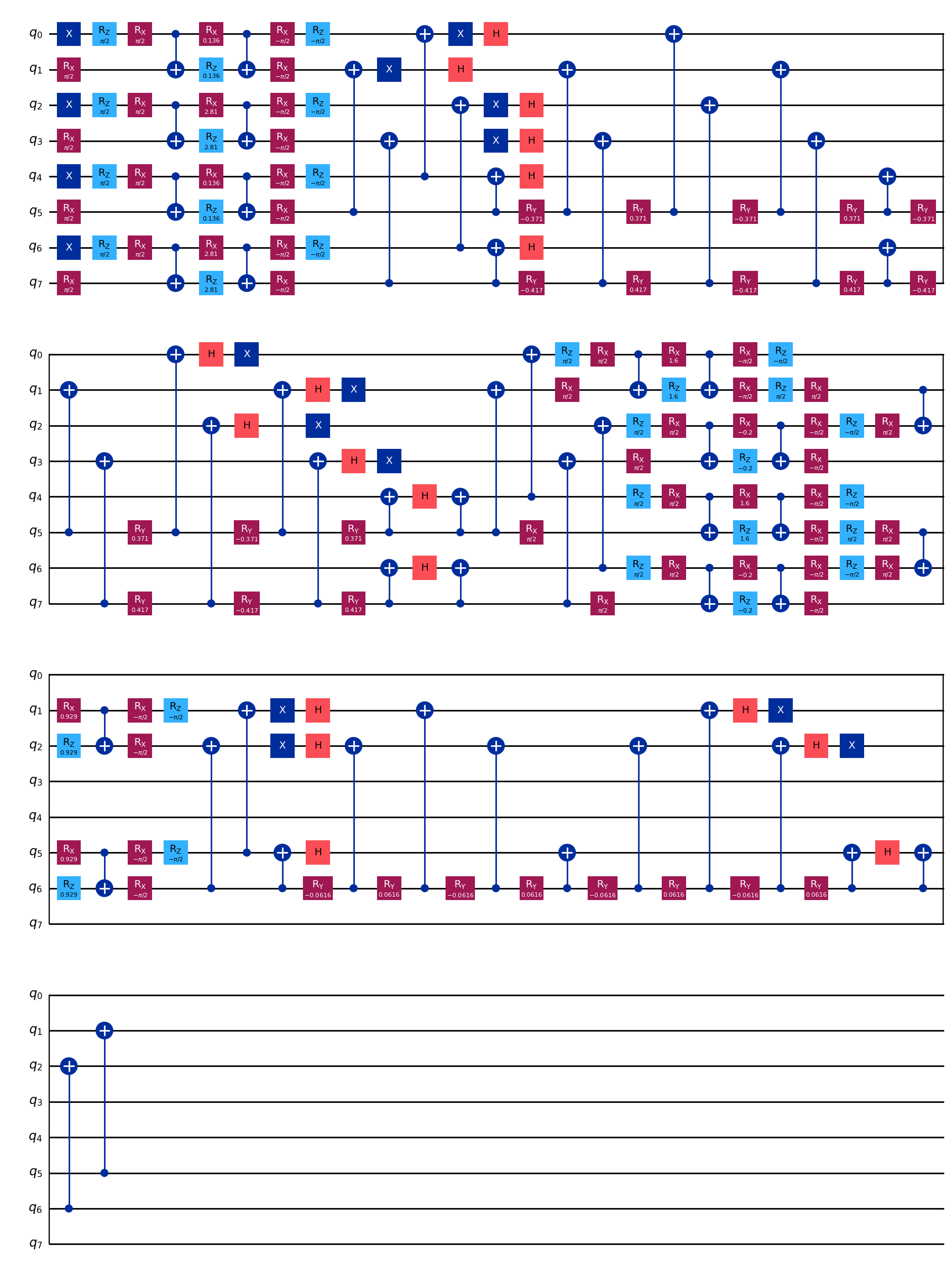}
    \caption{Circuit of the one layer pp-tUPS using an active space (4,4).}
    \label{fig:pptUPS(1)_AS(4,4)}
\end{figure}

\begin{figure}[H]
    \centering
    \includegraphics[width=0.8\linewidth]{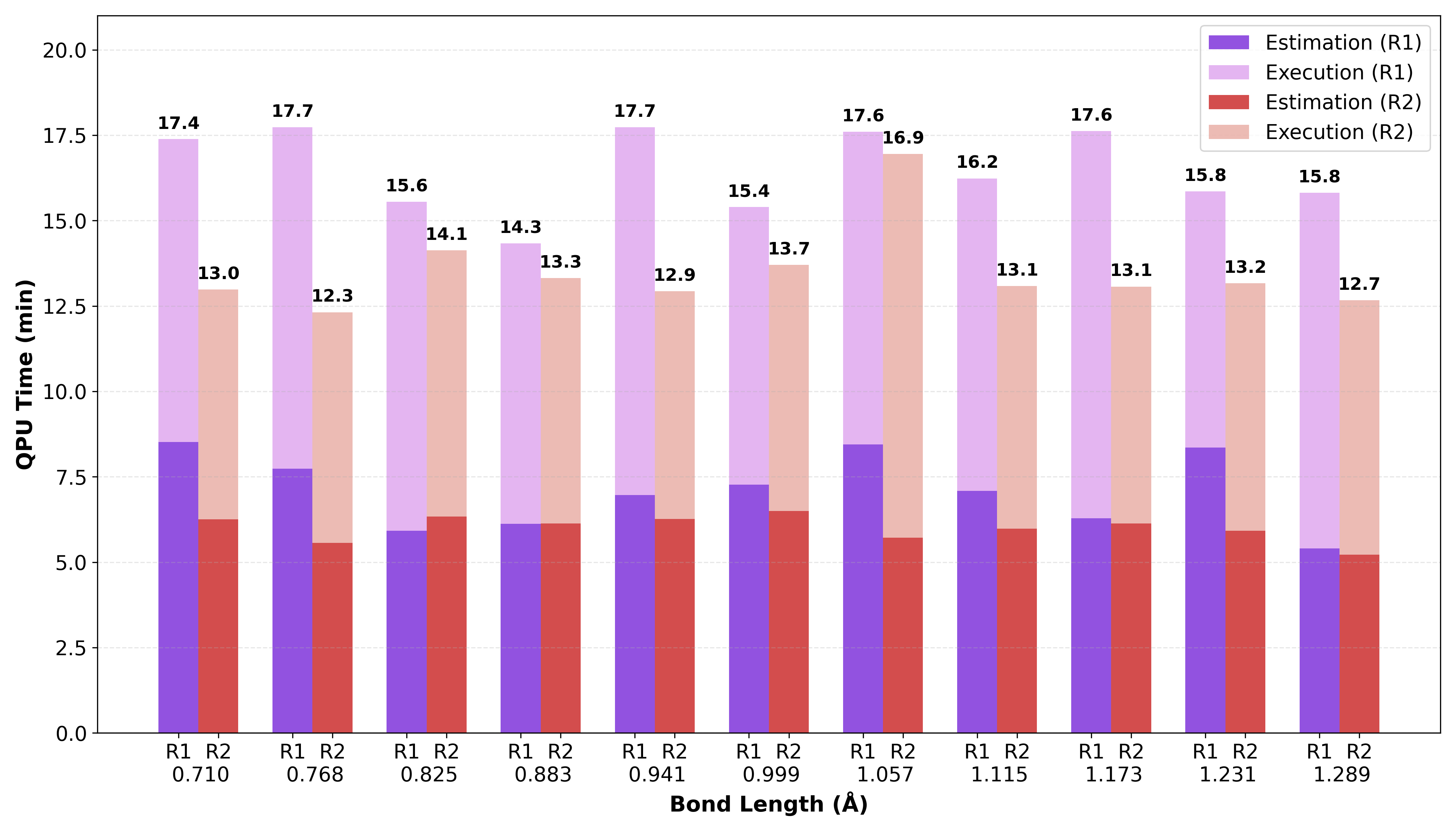}
    \caption{QPU time used for R1 and R2, and each geometry, with a 0.1 precision.}
    \label{fig:QPUTime_P01}
\end{figure}

\begin{figure}[H]
    \centering
    \includegraphics[width=0.8\linewidth]{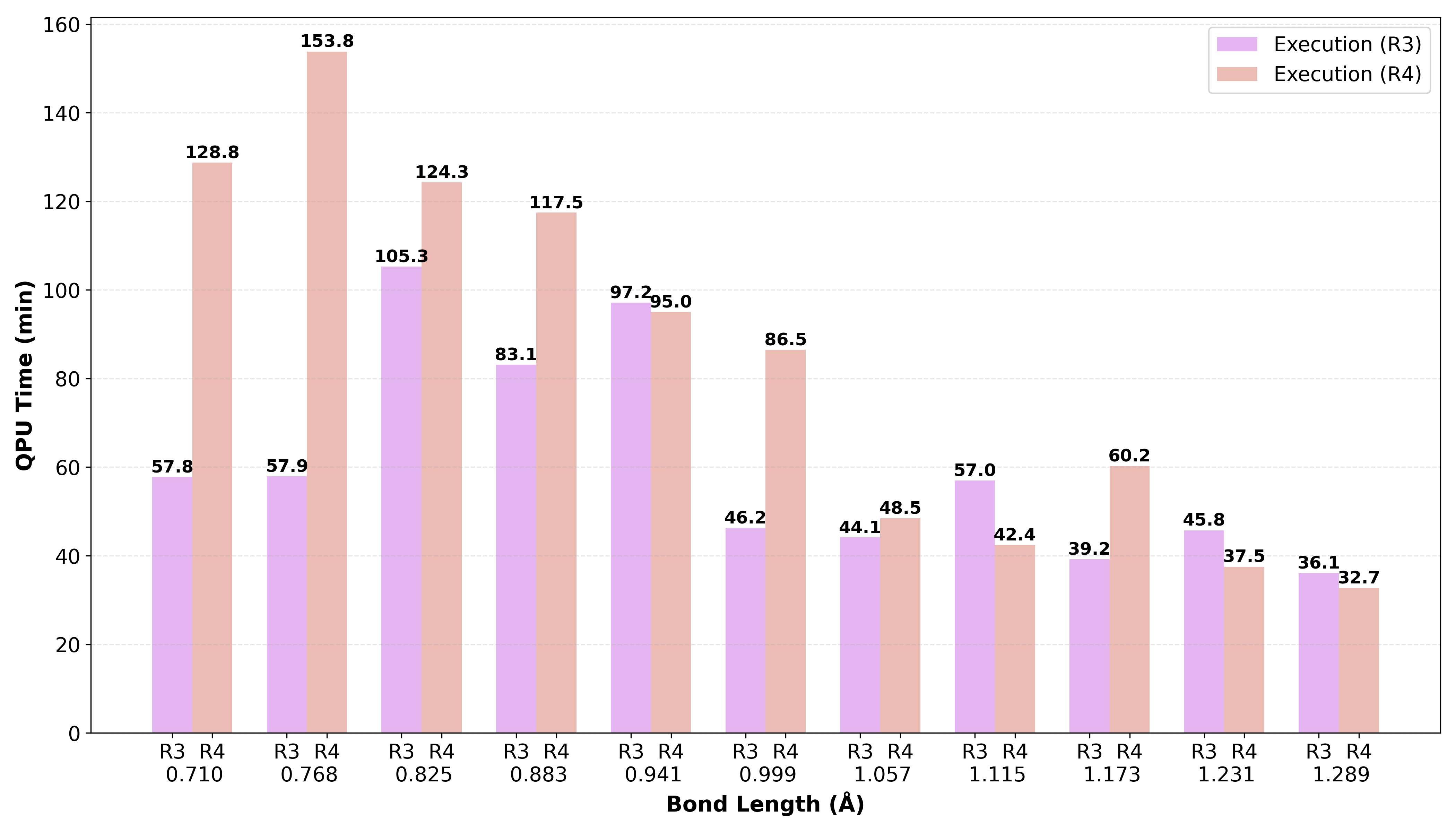}
    \caption{QPU time used for R3 and R4, and each geometry, with a 0.01 precision.}
    \label{fig:QPUTime_P001}
\end{figure}

\clearpage
\printbibliography


\end{document}